\documentclass[12pt]{article}
\usepackage{psfig,epsf}
\usepackage{a4wide,graphicx}
\usepackage{cite}

\def\NPA{{\em Nucl. Phys.} {\bf A}}
\def\NPB{{\em Nucl. Phys.} {\bf B}}
\def\PLB{{\em Phys. Lett.} {\bf B}}
\def\PREP{{\em Phys. Rep.}\ }

\def\PRL{\em Phys. Rev. Lett.\ }
\def\PRD{{\em Phys. Rev.} {\bf D}}
\def\PRC{{\em Phys. Rev.} {\bf C}}

\def\ANN{{\em Ann. Phys.}\ }

 \def\tstrut{\vrule height2.5ex depth0pt width0pt} 

\begin{document}
\title{
~ \hfill
~Non-localities and Fermi motion corrections in $K^-$ atoms}

\author{C. Garc\'{\i}a-Recio$^{a)}$, J. Nieves$^{a)}$, \\
E. Oset$^{b)}$ and  A. Ramos$^{c)}$ \\
a) Departamento de F\'{\i}sica Moderna,\\ Universidad de Granada,
 E-18071 Granada, Spain\\
b) Departamento de F\'{\i}sica Te\'orica and IFIC, \\Centro Mixto
Universidad de Valencia-CSIC, \\Institutos de Investigaci\'on de
Paterna, \\Aptdo. Correos 22085, 46071 Valencia, Spain\\ 
c) Departament
d'Estructura i Constituents de
la Mat\`eria, \\Universitat de Barcelona, 08028 Barcelona, Spain \\
}
\maketitle

\begin{abstract}
  We evaluate the p--wave $K^-N$ amplitudes from the chiral Lagrangians and from
  there construct the p--wave part of the $K^-$ nucleus optical potential plus a
  small s--wave part induced from the elementary p--wave amplitude and the nuclear
  Fermi motion. Simultaneously, the momentum and energy dependence of the s--wave
  optical potential, previously developed, are taken into account and shown to
   generate a small p--wave correction to the optical potential. All the
   corrections considered are small compared to the leading s--wave potential, and
   lead to changes in the shifts and widths which are smaller than the
   experimental errors.
      A thorough study of the threshold region and  low densities is conducted,
   revealing mathematical problems for which a physical solution is given.        

\end{abstract}
 PACS: 13.75.Jz; 21.30.Fe; 21.65.+f; 25.80.Nv; 36.10.Gv 

 [Key Words] Kaonic atom, Non-localities in optical potential, Chiral
 unitary model 

\newpage
\section{Introduction}

 The problem of kaonic atoms has attracted considerable attention along 
 the years \cite{alberg76,wei,miz,bat,Batty97,Friedman94,Friedman99,zaki,baca}
 and has regained interest recently due to the new perspective that the use of
 chiral Lagrangians has brought to the problem of the kaon interaction with a
 nuclear medium \cite{koc,waas,lut,ram}.  The need to obtain accurate results
 for the kaon selfenergy in a nuclear medium in view of the possibility to get
 kaon condensates in neutron-proton stars \cite{kaplan,gerry,pons} has also 
 added a renewed interest in the subject. Similarly, the interpretation
 \cite{cass,li} of the
 enhancement of the $K^-$ yields in heavy ion reactions \cite{barth,laue}
  relies on the value of
 the $K^-$ selfenergy in the nuclear medium. One of the characteristics of the
 $\bar{K} N$ interaction at low energies is the strong dominance of the s--wave
 amplitude, and the $\bar{K} N$ cross section to different channels can be very
 well described with only s--waves up to momenta of the order of 200~MeV or
 more \cite{martin,siegel,kai,ose,caro}. 
 
  The dominance of the s--wave in the
 elementary interaction  has been the justification for using 
 traditionally s--wave $\bar{K}$  
 nucleus optical potentials 
 \cite{alberg76,wei,miz,bat,Batty97,Friedman94,Friedman99,zaki,baca}, by means
 of which good agreement with data can be obtained.  This is in contrast with
 the situation in pionic atoms, since the relative small contribution of 
 the s--wave part of the potential, together with a sizeable p--wave part,
make
 this latter contribution quite important in the interpretation of the pionic 
 atom data \cite{eric,seki,juan}. 
 
    Theoretical evaluations of the optical potential for a particle in a
    nucleus usually start from the
  impulse approximation, or $t \rho$ form of the potential, with $t$ the
  elementary  scattering  amplitude of the particle with the nucleons averaged
  over isospin and the Fermi motion of the nucleons. Yet, it is known that
  the explicit consideration of the Fermi motion leads to corrections to this
  result. Certainly there are other higher order corrections in the density
  from mechanisms involving many nucleons, which one can systematically 
  tackle using many body techniques \cite{juan}, but even at the level of one
  single scattering the explicit consideration of Fermi motion brings
  corrections to the optical potential beyond the $t \rho$ approach. A detailed
  study of these correction terms for the case of pionic atoms was done in
\cite{carmen},
  where it was found that the s--wave elementary amplitude induces a p--wave
  term in the optical potential and, similarly, the p--wave elementary amplitude
  induces an s--wave term.   Yet, these corrections are small compared to the
  original impulse approximation and even small compared to the $\rho^2$ terms
  which account for pion absorption. 
  
     The aims of the present paper are to investigate the nonlocal
     effects induced form the s--wave part of the $K^-$ nucleus optical
     potential and also 
  to evaluate the p--wave part of it. Firstly, 
we begin from  the part of the potential evaluated  from the s--wave 
   $K^- N$ interaction in
  \cite{ram} and derive from there the nonlocal corrections originated from the
  $\omega$ and $k$ dependence of this potential. These non-local terms
  were evaluated in
  \cite{flor} and there they were found to be large and quantitatively important in the
     analysis of the spectra of kaonic atoms. We also evaluate the part of
  the potential originating from the elementary p--wave  $K^- N$ interaction and
  deduce from there the p--wave term of the optical potential plus an induced
  s--wave part due to the Fermi motion of the nucleons. 
  
  We shall see that all
  these corrections are rather small and do not appreciably change the results
  obtained from the s--wave part of the potential alone.  

  One important finding
  is that, in order to properly evaluate these corrections, one must pay
  a special attention to the region of low densities. Consequently we discuss 
  in detail the problems that one faces implementing the
low density limit around a threshold, particularly in the derivatives
of the optical potential.    
       
Finally, we study the effects of the inclusion of the $\Sigma^* -h$ excitation
     in the self-consistent calculation of the s--wave $K^-$
     potential.
We find that, though best-fits of better quality can be obtained, this
     new theoretical potential leads to an overall acceptable
     description of the measured shifts and widths.

\section{Brief summary of the $\bar{K} N$ and $\bar{K}$ nucleus interactions}

  For the elementary $\bar{K} N$ interaction we follow the 
chiral unitary model of \cite{ose}.  This work follows closely the steps
of \cite{kai}, where the usefulness of combining unitarity in coupled
channels with
the chiral Lagrangian was made manifest, but uses an enlarged basis of coupled
channels. 
The inclusion of all the channels formed by the octets of the pseudoscalar
mesons and stable baryons made it possible to obtain good solutions
in \cite{ose}
by means of only the lowest
order chiral Lagrangian and a suitable cut off to regularize the loop
integrals. On the other hand, in \cite{kai} the effect of the unopened channels
was accounted for by including higher order terms in the
chiral Lagrangian.

The $K^- N$ $t$ matrix, $t_{ij}$, is obtained in \cite{ose} through the
iteration of the  Lagrangian in a coupled channel
Bethe--Salpeter equation.

The
$K^-$ selfenergy is evaluated in detail in ref.\cite{ram} for
nuclear matter by means of the integral
\begin{equation}
\Pi(k,\rho)=2 \sum_N \int \frac{d^3p}{(2\pi)^3} n(\vec p, \rho)
~t_{K^-N}^{(m)} (k,\vec p,\rho),\label{eq:pik}
\end{equation}
where $N$ stands for protons or neutrons and $t^{(m)}$ is
the $K^- N$ scattering matrix in the nuclear medium with density $\rho$.
In Eq.~(\ref{eq:pik}),  $n(\vec p, \rho)$ denotes the occupation probability of momentum
states in the Fermi sea in the nuclear medium at finite density $\rho$.
Only
the s--wave amplitude is considered in this evaluation. In section 5 we shall
work out the contribution from the p--wave interaction.

The $t^{(m)}$ matrix is evaluated from the Bethe--Salpeter equation, but
modifications are done in the meson and baryon propagators of
the loops to incorporate the medium effects.  The states allowed
in the loops are $\bar K N$, $\pi \Sigma$, $\pi \Lambda$,
$\eta \Sigma$, $\eta \Lambda$, $K \Xi$.

The medium modifications implemented are the following:\\
\noindent
1) Pauli blocking in the nucleon propagators.  This effect was
proved to be very
important in \cite{koc,waas}.  Indeed, forcing the
intermediate nucleon states to be on top of a Fermi sea costs more
energy, and the net effect is a shift to higher energies
of both the real and imaginary parts
of the $K^- p$ amplitude which is dominated by the $\Lambda(1405)$ resonance
below threshold.
The shift of the real part automatically produces an attractive $K^-$
selfenergy already at very small densities.\\
\noindent
2) However, if the $K^-$ selfenergy acquires a negative value
then it costs less energy to produce the $\Lambda(1405)$ resonance
hence producing a shift of the $K^-p$ amplitude toward lower energies.
A selfconsistent evaluation becomes then necessary
as shown in  \cite{lut}, where it was found
that the consideration of the $K^-$ selfenergy together with
Pauli blocking on the nucleons left the position of the $\Lambda(1405)$
resonance basically unchanged.  In  \cite{ram} the $K^-$
selfenergy is also considered and a selfconsistent evaluation is also done.
\\
\noindent
3) In addition to the former ingredients new effects are considered in
\cite{ram}, i.e. the pion selfenergy in the $\pi \Sigma$, $\pi \Lambda$
channels is also taken into account allowing the pions to excite $ph$,
$\Delta h$ and $2p2h$ components.  Furthermore, the difference of binding
 between the nucleons or $\Sigma$ and $\Lambda$ hyperons is also incorporated.

The results obtained are qualitatively similar to those
found in \cite{lut} except that the imaginary part of the $K^- p$ amplitude
becomes even wider and essentially flattens at full nuclear density
$\rho \sim \rho_0$.

\section{Nonlocal terms associated to the s--wave part of the potential}

  As described in the former section the optical potential for the
  $K^-$ nucleus interaction was evaluated in nuclear matter as a
  function of the density in \cite{ram} and, in order to apply it to
  finite nuclei, the local density approach was used in \cite{zaki},
  something justified for the s--wave potential as discussed in
  \cite{juan}.  The $K^-$ selfenergy obtained in \cite{ram} from the
  interaction of the $K^-$ with protons and neutrons in symmetric
  nuclear matter has an explicit dependence on $k^0=\omega$ and $\vec
  k$, the energy and momentum of the antikaon.  However, in order to
  solve the Klein--Gordon equation (KGE) to obtain energies and widths
  of the kaonic atoms in \cite{zaki}, the potential was evaluated at
  the $K^-$ threshold ($\omega = m_K , \vec k = 0$).  In what follows
  we derive the corrections to the optical potential from the
  consideration of the explicit $\omega$ and $\vec k $ dependence of the
  kaon selfenergy in the nuclear medium.

 We write the $K^-$ selfenergy in nuclear matter, $\Pi$, as
\begin{eqnarray}
\Pi(\omega,\vec{k},\rho) = 2\omega V_{opt}&=& \Pi(m_K,0,\rho) +
     b(\rho) ~\vec{k}\,^2 +
     c(\rho) ~(\omega - m_K) 
\ ,\label{eq:pibc}
\end{eqnarray}
with 
\begin{eqnarray}
b(\rho) &=&\frac{\partial\Pi}{\partial
 \vec{k}\,^2}\Bigg|_{(\omega=m_K,~\vec k = 0)}
\ , \quad
 c(\rho) = \frac{\partial\Pi}{\partial \omega}\Bigg|_{(\omega=m_K,~\vec k
= 0)}
 \ ,\label{eq:bc}
\end{eqnarray}
 where the second order corrections in ${\vec k}\,^2$
 and $(\omega-m_K)$ have been neglected.
   
 Once at this point the momentum $\vec k$ is not defined for the bound
 $K^-$ in the atom and
 instead it becomes an operator. In detailed studies of finite nuclei one can
 trace the origin of this operator and how it acts on the density-dependent
functions of the
 potential or the kaonic wave function. For instance, in \cite{juan} one can
 see that the $\vec{k}\,^2$ which appears in the p--wave part of the
 $\pi$-nucleus optical potential
 evaluated in nuclear matter corresponds in the equivalent finite nucleus
 calculation to the combination 
\begin{equation}
\vec{k}\,^2_{CM} f(\rho) \rightarrow \frac{1}{(1+\varepsilon)^2}
\left[ -\vec{\nabla} f(\rho(\vec{r}\,)) \vec{\nabla} + \frac{1}{2}
\varepsilon \vec{\nabla}\,^2 f(\rho(\vec{r}\,)) \right] \ ,\ 
\varepsilon={m_\pi\over M_N}\ .
\end{equation}

     In the present case the evaluation of the $K^-$ selfenergy in finite
 nuclei, with all the effects considered in \cite{lut,ram} plus the requirement
 of selfconsistency is a rather involved task, which would become
 advisable should these nonlocal effects be too big.  Yet, as we shall see, the
 effects are small, smaller than present experimental uncertainties in the data,
 and thus the estimates which we shall perform here are sufficient to establish
 the relevance of these effects.  One of the handicaps of having evaluated the
 selfenergy in infinite nuclear matter is that we do not know to which kind of
 $\vec{\nabla}\vec{\nabla}$ operator will the factor  $\vec{k}\,^2$
correspond to. In order
 to estimate the size of the corrections and the uncertainties, we shall work
 with some different assumptions which are guided by the results obtained for
 pionic atoms in the translation from infinite matter to finite nuclei.
  The types of operators used for the $b ~\vec{k}\,^2$ term of the
 $K^{-}$-selfenergy in  
Eq.~(\ref{eq:pibc}) are shown below
\begin{eqnarray}
{\rm a)}~ b ~\vec{k}\,^2 & \to & -\vec{\nabla} b \vec{\nabla} \nonumber \\
{\rm b)}~ b ~\vec{k}\,^2 & \to & -\vec{\nabla} b \vec{\nabla} - \frac{1}{2} (\Delta b) \\
{\rm c)}~ b ~\vec{k}\,^2 & \to & - b \vec{\nabla}\,^2 \nonumber
\end{eqnarray}  
  
  The form a) is the Kisslinger type of interaction $\vec{k}\cdot b
  \vec{k}$,  the form b) appears in
  \cite{carmen} by using the Wigner transform of the symmetrized form 
  $(b\vec{k}\,^2+\vec{k}\,^2b)/2$, the form c)
  allows the $\vec {k}^2$ operator to act directly on the kaonic wave function,
  and thus has a special physical significance.
  
    We can also adopt a different point of view and, since the results
    of Eq.~(\ref{eq:pik}) 
    already come from using local approximations implicit in the use of a
    local Fermi sea, we can convert the nonlocal potential of
$\vec{k}\,^2$
    into a local, but energy dependent, potential. The trade between
nonlocal
    and local energy dependent potentials is a technique often used in many
    body theory \cite{mahaux}. It is based on the use of the Schr\"odinger
    equation (KGE in our case).
    The KGE is written here as
    
\begin{equation}
[ - \vec{\nabla}\,^2 + \mu ^2 + \Pi (r)] \phi (\vec{r}\,) =
[\omega - V_{C} (r) ]^2 \phi (\vec{r}\, ) \ . \label{eq:kg}
\end{equation}
where $\mu$ is the kaon-nucleus reduced mass and $V_{C}(r)$ is the
Coulomb potential with a finite nuclear size and vacuum-polarization
corrections.

   In view of Eq.~(\ref{eq:kg}), and the fact that the meaning of
$\vec{k}\,^2$ 
   in Eq.~(\ref{eq:pibc}) is the kaon momentum squared, we can take for it
the expectation
   value of $-\vec{\nabla}^2$ for the kaonic wave functions. This leads to 
   
\begin{eqnarray}
{\rm d)}~b ~\vec{k}\,^2 & \to & b \left[ (\omega-V_C)^2 - \mu^2 - \Pi \right]
\label{eq:d}
\end{eqnarray}
   
   Also since the potential is complex we can take the more symmetrical
   situation $(b\vec{k}\,^2+\vec{k}\,^2b)/2$, in which case we obtain the 
   real part of the expression in Eq. (\ref{eq:d}) 

\begin{eqnarray}
{\rm e)}~ b ~\vec{k}\,^2 & \to & b \,{\rm Re\,}\left[ (\omega-V_C)^2 - \mu^2 - \Pi \right]
\end{eqnarray}
   
   The energies and shifts for the energy dependent potential can be obtained
   by iteration. Given the smallness of the pieces under consideration the
   convergence is extremely fast. 
   
\section{Threshold behaviour and the low density limit}

 Eqs.~(\ref{eq:bc}) require the evaluation of the derivatives of the
  kaon selfenergy with respect to $\vec{k}\,^2$ and $k^0$ at
threshold. However, 
  these magnitudes are problematic when one goes to small densities,
   as we shall see.  The problems stem from the behaviour of the
 free $t$ matrix at threshold which has a cusp.  This is related to the
 contribution of the elastic channel $K^-N \to K^-N$ and hence, for the
 discussion, one can neglect all the other channels.  Also the discussion here 
 is completely general and thus we do not particularize to the $K^-N \to K^-N$
 reaction.

\subsection{Behaviour of the elastic rescattering terms}
   
 To simplify the discussions we shall take unity for the transition
 potential $K^-N \to K^-N$. We begin with the study of the second
 order rescattering term with this potential, see
 Fig.~\ref{fig:diagram}. This is the term studied in the appendix of
 Ref.~\cite{flor}, which there led to coefficients $b$ and $c$
 behaving like $k_F^2$ in the limit of low densities, a result that
 would seem to violate the low density theorem.  
\begin{figure}[hbt]
\begin{center}
\includegraphics[width=8cm,angle=0]{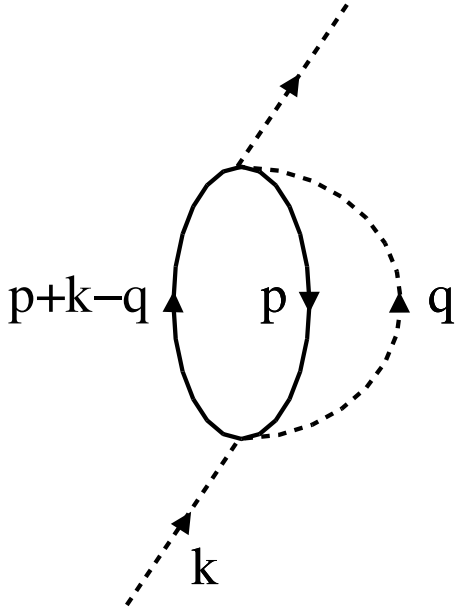}
\caption{Second order rescattering contribution to the meson selfenergy.}
\label{fig:diagram}
\end{center}
\end{figure}

    In order to clarify the problems let us write the kaon selfenergy
for this rescattering term 
\begin{equation}
\Pi^{res}(k^0,\vec{k\,},\rho)=4\int\frac{d^3p}{(2\pi)^3} \int \frac{d^3q}{(2\pi)^3}
\frac{1}{2\omega(q)} \frac{n(\vec{p}\,,\rho) [1-n(\vec{p}+\vec{k}-\vec{q}\,,\rho)]}{k^0-
\omega(q) +\varepsilon(\vec{p}\,)-\varepsilon(\vec{p}+\vec{k}-\vec{q}\,)+i\eta} \ ,
\label{eq:rescat1}
\end{equation} 
where we use relativistic energies for the meson  but non-relativistic ones for the
nucleon, in  order to be able to use explicit formulae for the Lindhard functions. 
In the following, we will denote the meson and nucleon masses by $m$ and $M$,
respectively. We have
 written explicitly the integral over the Fermi sea momentum, $\vec{p}$, and over
 the meson momentum, $\vec{q}$.  Let us examine the behaviour of the
$\vec{q}$ integral in Eq.~(\ref{eq:rescat1}) in the limit of zero Fermi momentum and
$\vec{k}=0$.  In such a case one gets 
\begin{equation}
t^{res}(k^0,\vec{k\,}=0)=\frac{4\pi}{(2\pi)^3}\int q^2 dq \frac{1}{2\omega(q)}
\frac{1}{k^0-\omega(q) -\frac{q^2}{2M} + i\eta} \ , 
\label{eq:rescat2}
\end{equation} 
 which would give us $t^{res}$, which is the contribution of the rescattering term to the free $t$ 
matrix,   
 as a function of $k^0$.   Now let us
 find out the derivative of this function with respect to $k^0$ and
 particularize for $k^0=m$. Commuting the partial derivative and the integral,
 and concentrating in the region of small $q$, one finds an infrared divergence of
 the type 
\begin{equation}
\int dq \frac{1}{q^2}
\label{eq:infrared}
\end{equation}

 Certainly, as one puts a finite, even if small, Fermi momentum, the
 divergence disappears, and one finds a $1/k_F$ behaviour, similarly
 to the derivative of the function $I(\omega,0)$ with respect to
 $\omega$ in the appendix of \cite{flor}. Obviously, the divergence is
 recovered in the strict limit of $k_F=0$.  This is a  subtle
 problem.  One reason is that we have a branch point at threshold
 and the commutation of the derivative and the integral made above is
 not justified. Indeed, one can do the integration with respect to
 $q$, in Eq.~(\ref{eq:rescat2}), analytically and, after
 renormalization of the ultraviolet behaviour, the integral can be
 obtained from the two particle loop integral, $J_0$, given 
 in appendix A of~\cite{na99}. The difference is that in
 Eq.~(\ref{eq:rescat2}) we use a non-relativistic propagator for the
 nucleon, but since we are concerned about the infrared divergence it
 does not make any difference. Then one can study the derivative with
 respect to $k^0$. 
 The expression for the loop function is:

\begin{eqnarray}
t^{res}(k)&\simeq&2MJ_0(k^2)
\equiv 2M~i\int\frac{d^4q}{(2\pi)^4}~\frac{1}{(q^2-M^2+i\eta)}~\frac{1}{((q+k)^2-m^2+i\eta)},\nonumber
\end{eqnarray}
where the approximation sign is present because the l.h.s. of the
equation is the baryon non-relativistic approach to the
r.h.s. expression. 
Removing constant subtraction terms
irrelevant for the discussion, this expression~\cite{na99} is given by
 \begin{eqnarray} (4\pi)^2 J_0(s)&=& 
 (1-\frac{(M-m)^2}{s}) ~\sigma~\ln\frac{\sigma +
 1}{\sigma -1} +\frac{M^2-m^2}{s} \ln\frac{M}{m}\ ,\nonumber \\ 
\sigma&=&\sqrt{\frac{1-\frac{(M+m)^2}{s}}{1-\frac{(M-m)^2}{s}}}\ ,
\label{eq:loop1}
\end{eqnarray}
 where  $s$ is the Mandelstam variable.
Close to threshold (in this case $s=(M+m)^2$), and above it, we can expand $J_0(s)$ as 
 \begin{equation}
J_0(s) =
\frac{1}{(4\pi)^2}
\left( 
(1-\frac{(M-m)^2}{s})(-i\pi\sigma + 2\sigma^2+{\cal O}(\sigma^4))
+\frac{M^2-m^2}{s} \ln\frac{M}{m}
\right) \ , 
\label{eq:loop2}
\end{equation}
 and then we can perform the derivative with respect to $s$ and find $\left(
 \frac{\partial \sigma}{\partial s} =
 \frac{2mM}{\sigma(s-(M-m)^2)^2} \right)$  
 at threshold from above:
 \begin{equation}
\frac{\partial J_0(s)}{\partial s} =  \frac{1}{(4\pi)^2}\frac{1}{(M+m)^2} 
\left( -i\frac{\pi}{2\sigma} + 2  - 
\frac{M-m}{M+m} \ln\frac{M}{m} + {\cal O}(\sigma) \right)
 \ .
\label{eq:loopder}
\end{equation}
 Thus, what we find is that, since $\sigma$ goes to zero at threshold, the
 derivative of the imaginary part goes to infinity but the derivative of the
 real part remains finite.  This is related to the fact that the imaginary part
 goes as the momentum of the particle, and then its derivative with respect to 
 $s$ is infinite at threshold. Below threshold $\sigma$ becomes
purely imaginary and therefore $\partial J_0(s)/ \partial s$ is purely
real and it diverges at threshold. 
 This is logical since the analytical
 continuation  below threshold of the imaginary part above threshold  becomes 
 real and this is the
 reason for the infinite derivative. Hence, the derivative of $J_0(s)$
with respect to $s$, and therefore that with respect to $k^0$ when
$\vec{k}=0$, takes different values at the right hand side than at the
 left hand side of the threshold point.

 The results are then different from what one obtains by commuting the
 integral and the derivative, as we did before in
 Eq.~(\ref{eq:rescat2}) to obtain the divergence of
 Eq.~(\ref{eq:infrared}).  However, the fact remains that there are
 divergences. Nevertheless, this analytical study has served to see
 that the origin of the divergence is the existence of the imaginary
 part in the free $t$--matrix close and above threshold.  This
 realization is important because, as we shall see, for finite values
 of $\rho$, the effect of
 Pauli blocking drastically reduces the imaginary part of the
 rescattering term of the meson selfenergy close to threshold, 
and what is more important 
it can be differentiated with respect to
 $k^0$. In order to see this, we perform first the integration over the
 Fermi sea in Eq.~(\ref{eq:rescat1}) and write the meson selfenergy as
 \begin{equation} \Pi^{res}(k^0,\vec{k\,}=0,\rho)=\int \frac{d^3q}{(2\pi)^3}
 \frac{1}{2\omega(q)} U(k^0-\omega(q),q,\rho) \ ,
\end{equation}
 where $U$ is the Lindhard function for forward going particle-hole excitation. 
 The ordinary Lindhard function contains also the backward going particle-hole term.
 For this reason we give in the appendix the explicit expressions for the forward going
contribution to the Lindhard function needed here. 
 Inspection of  Eq.~(\ref{eq:rescat1}) tells us that there is only imaginary part for
$k^0-\omega (q)>0$.

 It is easy to prove, using the formula for Im$\,U$ of the appendix, that, for 
 $k^0-m<<k_F^2/2M$, ${\rm Im\,}\Pi^{res} (k^0,0,\rho)$ goes as
 $(k^0-m)^2$, but we save the proof here since this is a well known result for
 the imaginary part of the selfenergy of any particle in a medium
 \cite{mattuck,walecka,brown}. This result is important because then the
 imaginary part of the meson selfenergy and its derivative with
respect to $k^0$ are continuous 
  at threshold (actually for this rescattering contribution both the
imaginary part of the function and its derivative vanish at
threshold). Hence, the singularity in the derivative of the imaginary
part of the free $t$ matrix above threshold, which we have seen in
Eq.~(\ref{eq:loopder}), disappears in the meson selfenergy at finite densities. The
use of dispersion relations in the selfenergy and its derivative then
guarantee that both the real part of the selfenergy and its derivative
are also continuous at threshold.  Let us show this in a quantitative manner.

\begin{figure} 
\begin{center} 
\includegraphics[width=10cm,angle=-90]{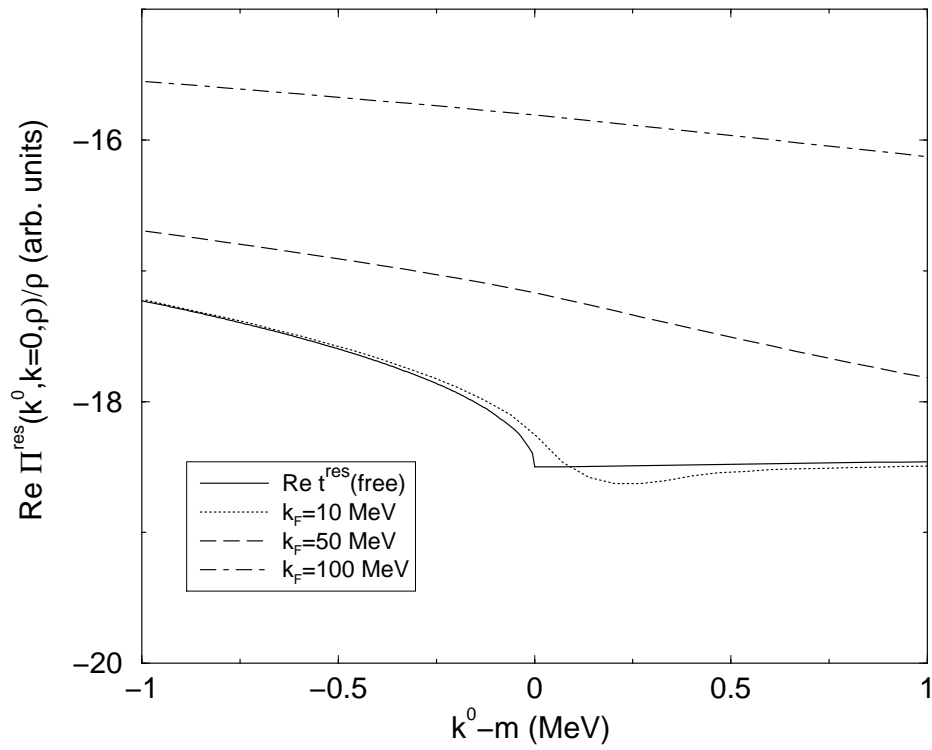}
\caption{Real part of the kaon rescattering selfenergy, $\Pi^{res},$
of Eq.~(\ref{eq:rescat1}) divided by $\rho$ as a function of $k^0-m$
for different densities.  }
\label{fig:pik0} 
\end{center}
\end{figure}

   In Fig.~\ref{fig:pik0} we show the real part of the meson
selfenergy from this rescattering piece for $k_F= 100$ MeV,
corresponding to $\rho=0.05 \rho_0$, 50 MeV ($\rho=0.006 \rho_0$) and
10 MeV ($\rho=0.00005 \rho_0$) divided by $\rho$ for normalization,
and we compare the results with the real part of the free $t$ matrix
for the meson nucleon interaction from the corresponding rescattering
term, $t^{res}$.  We observe, indeed, that for finite densities the
meson selfenergy is a regular function and shows no cusp, which is
clearly visible in the free $t^{res}$ matrix. It is interesting to see that
for very small densities (see the curve corresponding to $k_F=10$ MeV
in the figure) $\Pi^{res}/\rho$ goes like the free $t^{res}$ matrix, except for
the fact that Pauli blocking has provided a regularization 
around the cusp. It is also interesting to observe that even at
$k_F= 50$ MeV, and certainly at $k_F= 100$ MeV, one obtains
practically a linear function in $k^0$. We can thus anticipate that
problems linked to the original discontinuity of the derivative in the
$t$ matrix will have negligible effects in the evaluation of any
observable where the different densities will have to be weighed in
the nucleus.  However, the formal problem still remains, because one
has to define the derivatives at threshold for any density and, even
if now they can be calculated, for very small densities (see e.g. the
case of $k_F=10$ MeV) the function around threshold is by no means
linear in $k^0$. Hence, if one takes a linear extrapolation based on
the derivative at threshold, see Eq.~(\ref{eq:pibc}), it is obvious
from the figure that it would lead to a strong diversion from the
actual calculated values of the selfenergy.
  
     As we have mentioned, the problem is a formal one affecting only
to small 
     densities which should have no relevance in practical problems provided a
     sensible and realistic approach is taken. Yet, this might not be always
     easy to achieve. As an example we could consider a sensible approach the one
     in \cite{flor}, where, according to the authors, they "artificially modify
     the slope of the function such that it is zero at $k_F=0$ but agrees to
     good accuracy with $b_{eff}(k_F)$ for $k_F>$ 50 MeV ". 
However, the final effect of the nonlocalities on the calculated
level shifts and widths in that work is substantially larger than
we find here when treating carefully the nucleus as a finite system,
as described below.

    Indeed, the problems mentioned above do not appear in a finite
nucleus.   This issue was already
  addressed in \cite{jesus} in a different problem, the study of the effect of
  core polarization on inelastic or charge exchange pion nucleus reactions.
  The reason is simple but subtle. If one looks at the expression of the 
  Lindhard
  function, which results from removing the integral over $d^3q/(2\pi)^3$, the
  factor $1/2\omega (q)$ and setting $q=0$ in Eq.~(\ref{eq:rescat1}), and one takes
the threshold 
  value $k^0=m$ and
  $k=0$, one realizes that there is an indetermination of the type $0/0$ in the
  integrand.  The analytical expression of the Lindhard function is such that 
  its limit for $k\to
  0$ is finite. Yet, in a closed shell nucleus the numerator of the
  equivalent response function 
  is zero because it involves a matrix element $<\phi_1| {\rm exp}
  (i\vec{k}\vec{r}\,)|\phi_2>$, 
  where $\phi_1$ would correspond to an occupied state and $\phi_2$ to an
  excited state and these
  functions are orthogonal. However, the denominator would involve
  $\epsilon_i-\epsilon_j$, which is strictly non zero because
  there is a minimum excitation energy from the occupied states to the excited
  states.  Thus, the corresponding response function in a finite nucleus is
  strictly zero while the Lindhard function does not vanish because of  
  the continuity of
  the  energies in the Fermi sea in infinite nuclear matter.
   Rather
  than doing an unnecessarily complicated evaluation in a finite nucleus,
   a simple solution
  to the problem was given in \cite{jesus}, including a finite excitation
  energy in the particle state, which we shall call the gap, $\Delta$, and reevaluating
   the Lindhard function, which then
  turned out to be  strictly zero in the limit of small $\vec{k}$.  The expressions for the
  Lindhard function with the gap can be found in the appendix of \cite{jesus},
  however they include forward and backward propagating $ph$ excitations. We need
  here only the forward propagating $ph$ excitation according to
Fig.~\ref{fig:diagram}, and since
  we perform evaluations with this new function in what follows, we give the
  explicit expression for this function in the appendix. Once the gap
is included, the rescattering term of the meson selfenergy, $\Pi^{res}$, has
imaginary part only for $k^0 > m +\Delta$ and, thus, around
$k^0=m$, the relevant region for kaonic atoms, it is purely real. The gap is physical
  and it is about 1-3 MeV for most of the nuclei we analyze here. So, our
  strategy will be, first to show that once the gap is included the
selfenergy can be accurately approximated by a linear function of
$k^0$ around $k^0=m$. Second, to test that the results for kaonic atoms are
rather insensitive to the gap value, with changes on the shifts and
widths far smaller  than the experimental uncertainties.
  
\begin{figure}
\begin{center} 
\includegraphics[width=10cm,angle=-90]{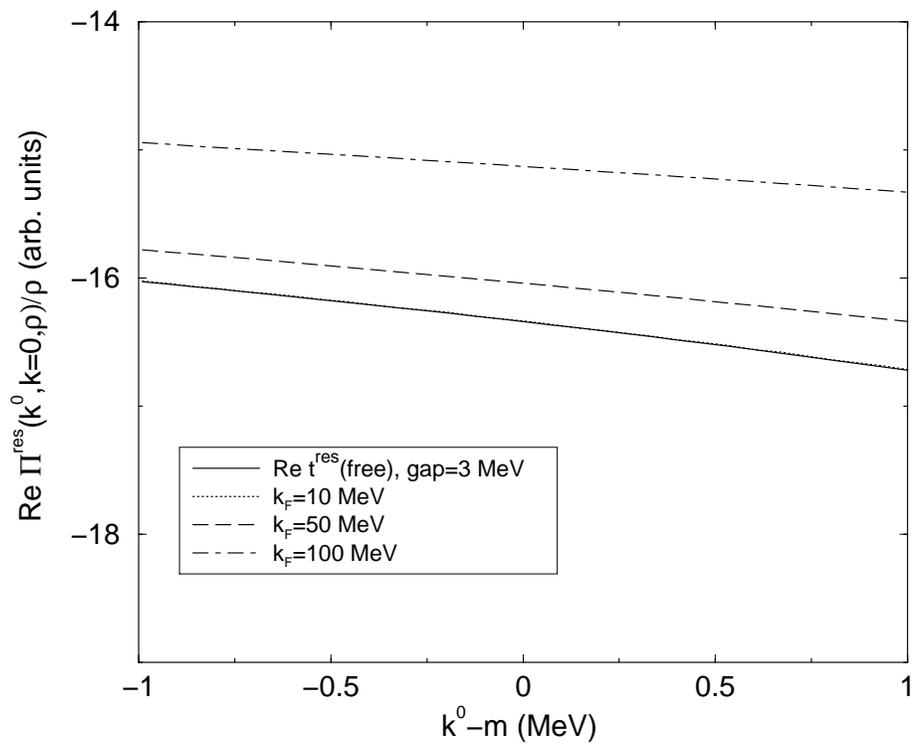}
\caption{Same as Fig.~\ref{fig:pik0}, but including a gap of 3 MeV in
the nucleon particle energy.}
\label{fig:pik0g} 
\end{center}
\end{figure}

   In Fig.~\ref{fig:pik0g} we show the results for the real part of
the rescattering term of the meson selfenergy divided by $\rho$ for
different densities and compare them with the real part of the free
$t^{res}$ matrix including the gap in the particle energy, this is to
say, we increase the nucleon mass in the intermediate state of the
rescattering term by the gap energy, i.e, we add $(-\Delta)$ in the
denominator of Eq.~(\ref{eq:rescat2}). We have chosen here a gap of 3
MeV.  The function around threshold is now well behaved for any
density, showing a linear dependence in $k^0$, and we can see that for
values of $k_F$ around 50 MeV the meson selfenergy divided by $\rho$
and the free $t^{res}$ matrix differ in less than two percent and they
are practically indistinguishable at $k_F=10$ MeV, a consequence of
the low density theorem for the meson selfenergy
\cite{dover,hufner,hufner2}.

      We did not yet pay attention to  the dependence of the selfenergy on
$\vec{k}\,^2$,
       but the same arguments as above 
    can be repeated. Again we can see in Fig.~\ref{fig:pik2g} that, once the gap
energy is     considered, the dependence
   is smooth around threshold, $\vec{k}\,^2=0$, and  the selfenergy divided
by $\rho$ merges 
  to the free $t^{res}$ matrix as a necessary consequence of the low
density theorem. Here $t^{res} (k^0=m,k)$ is defined as in Eq.~(\ref{eq:rescat2}) by
changing $\vec{q}\,^2/2M$ by $(\vec{k}-\vec{q})^2/2M$ (and adding
$(-\Delta)$ in the denominator).

{   
\begin{figure}
\begin{center} 
\includegraphics[width=10cm,angle=-90]{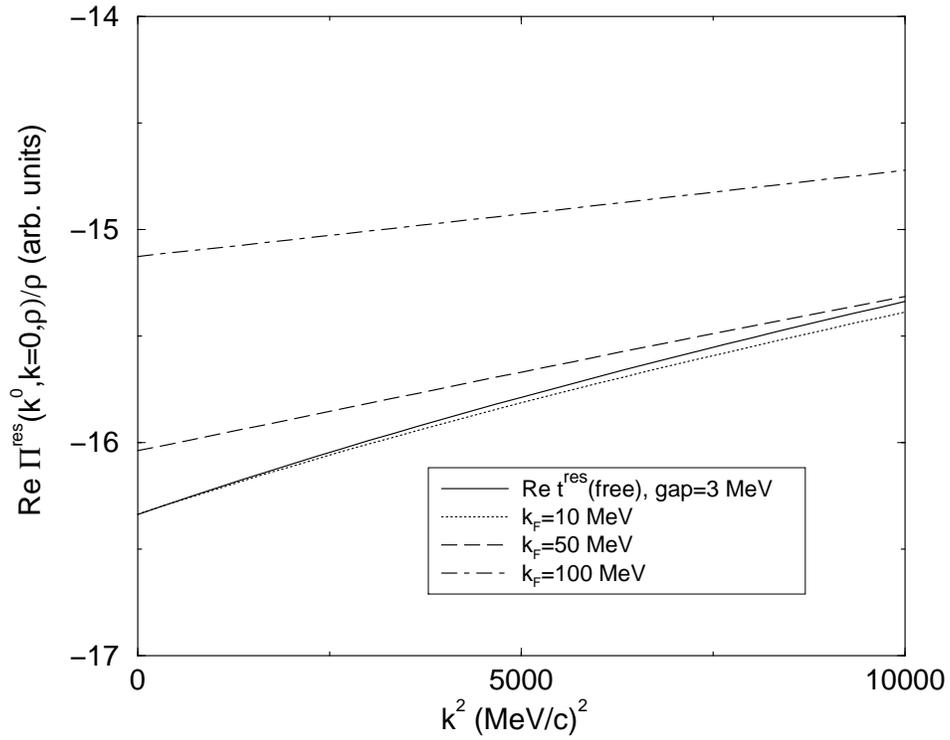}
\caption{Real part of the rescattering term contribution to the meson
selfenergy, $\Pi^{res}$, divided by $\rho$ as a function of $k^2$ for different
densities, evaluated including a gap of 3 MeV in the nucleon particle
energy.  }
\label{fig:pik2g} 
\end{center}
\end{figure}           
\newpage}

       The study conducted above has served to show the problems in the low
   density limit close to threshold due to the cusp behaviour of the free $t$
   matrix, but we have also seen that, once a physical value of
   the gap is introduced, all the problems disappear and one can also rely on the
   low density limit in order to evaluate values of the selfenergy and their
   derivatives with respect to $k^0$ and $\vec{k}\,^2$ at low densities. 
   
\subsection{Behaviour of the full kaon selfenergy}

      Now we come back to the realistic situation of the kaon selfenergy including
   all coupled channels in the calculation. We have learned from the
previous section that the kaon selfenergy does not behave linearly in
$k^0$, $\vec{k}^2$ around thresholds. We have avoided these problems
around the $K^-N$ threshold by means of the gap $\Delta$. The rest of
the thresholds, like $\pi\Sigma$, $\eta\Lambda$, etc., are far from
the region of energy and momenta met by the $K^-$ atoms, and hence
these other coupled channels do not affect the linear behaviour of the
kaon selfenergy for kaonic atoms. We will obtain
the low 
   density limit simply taking the selfenergy equal to $t\rho$, 
where $t$ stands now for the average of $t_{K^- N}$ over protons and
neutrons evaluated including the gap energy
   for the intermediate nucleon state in the Bethe--Salpeter equation. Thus, in the limit of $\rho$
going to zero
   we obtain
\begin{eqnarray}
\frac{\partial \Pi(k,\rho)}{\partial \vec{k}\,^2}
\Bigg|_{(\omega=m_K,~\vec k
= 0)}  &=& - \left. \frac{\partial t}{\partial
s} \right|_{(0)} \rho  
\nonumber \\
\frac{\partial \Pi(k,\rho)}{\partial \omega} \Bigg|_{(\omega=m_K,~\vec k = 
0)} &=& \left. \frac{\partial
t}{\partial s} \right|_{(0)} 2(m_K+M_N) \rho
\ ,
\label{eq:lowrho1}
\end{eqnarray}
where the label $(0)$ indicates that the derivative is taken at threshold.
  This means that
   the values of $b(\rho)$ and $c(\rho)$ from Eqs.~(\ref{eq:bc}) 
    can be cast in the low density limit as
\begin{eqnarray}
     \left.\frac{b(\rho)}{\rho}\right|_{\rho=0} &=&
     -\left. \frac{\partial t}{\partial s}\right|_{(0)} \nonumber \\ 
     \left.\frac{c(\rho)}{\rho}\right|_{\rho=0} &=&
      \left. \frac{\partial t}{\partial
s}\right|_{(0)} 2 (m_K+M_N) \ ,
\label{eq:lowrho2}
\end{eqnarray}
and the above relationships are the constraints we impose to the low
density behaviour of the parameters $b(\rho)$ and $c(\rho)$. Note that
these coefficients are now complex because of the inclusion of the
non-elastic channels.

\begin{figure}[]
\begin{center}
\leavevmode
\epsfysize=650pt
\makebox[0cm]{\epsfbox{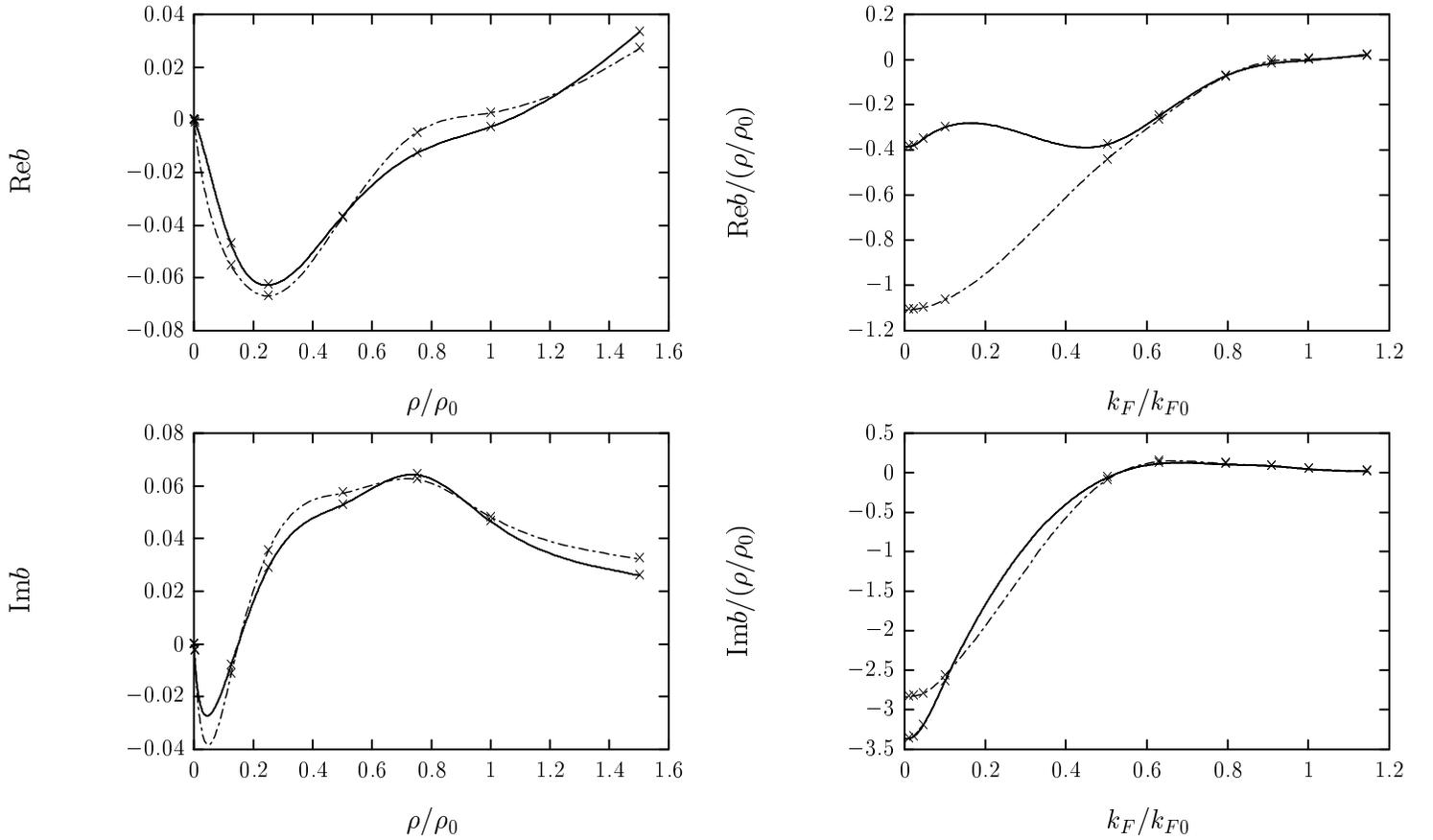}}
\end{center}
\vspace{-9.4cm}
\caption[pepe]{{ The real and imaginary parts of $b$ versus $\rho/\rho_0$ are depicted
in the two left figures. 
The real and imaginary parts of 
$[b/(\rho/\rho_0)]$ versus 
$k_F/{k_F}_0\equiv (\rho/\rho_0)^{1/3}$ are shown on the right panels.
The crosses
correspond to the points which have been numerically evaluated from
Eqs.~(\protect{\ref{eq:pik}}), (\protect{\ref{eq:bc}})  and 
(\protect\ref{eq:lowrho2}) and the
lines are the interpolated values, which have been used for
calculations.} The solid  lines have been obtained using an energy 
gap of $\Delta~=~1$~MeV and the dashed ones with $\Delta~=~3$~MeV.}
\label{fig:b}
\end{figure}

\begin{figure}[]
\begin{center}
\leavevmode
\epsfysize=650pt
\makebox[0cm]{\epsfbox{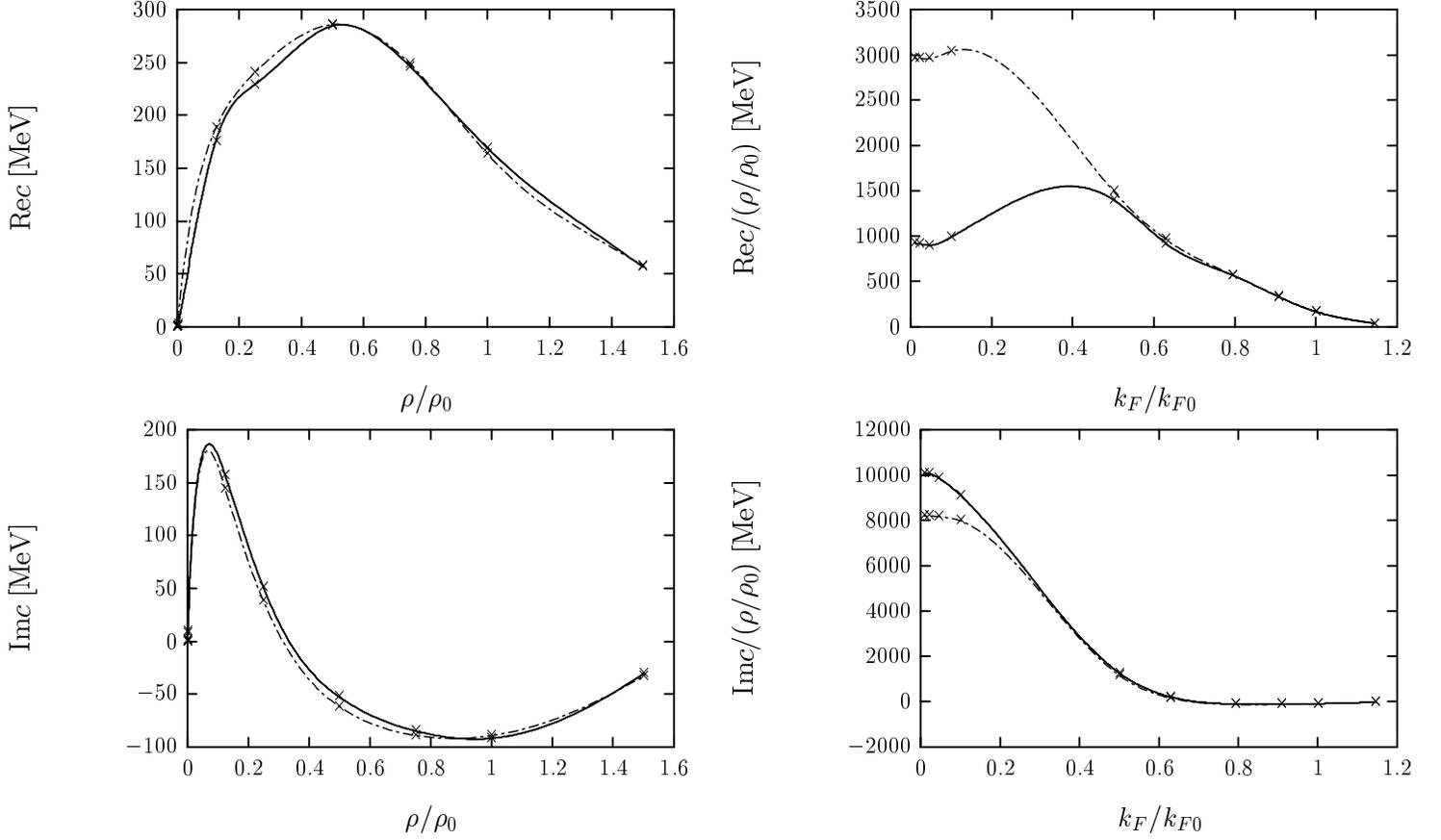}}
\end{center}
\vspace{-9.4cm}
\caption[pepe]{The left figures show the real and imaginary parts of $c$
versus $\rho/\rho_0$. The real and imaginary parts of 
$[c/(\rho/\rho_0)]$ versus 
$k_F/{k_F}_0\equiv (\rho/\rho_0)^{1/3}$ are shown in the 
figures on the right-hand side. The crosses
correspond to the points which have been numerically evaluated from
Eqs.~(\protect{\ref{eq:pik}}) , (\protect{\ref{eq:bc}}) and 
(\protect\ref{eq:lowrho2}) and the
lines are the interpolated values for arbitrary density. $c$ has been
obtained from $\Pi$ calculated without pion selfenergy for
$\rho/\rho_0\le 0.25 $.The solid  lines have been obtained using an energy 
gap of $\Delta~=~1$~MeV and the dashed ones with $\Delta~=~3$~MeV. }
\label{fig:c}
\end{figure}

   In Figs.~\ref{fig:b} and ~\ref{fig:c} we can see the numerical
   results obtained 
   for the partial derivatives of the kaon selfenergy as a function of
   the density. 
 In order to stress the low density behavior we have also shown them 
 divided by the
   density plotted against the Fermi momentum. The density and Fermi
   momentum are in units of $\rho_0=~0.17~$fm$^{-3}$ and ${k_F}_0~=~268.4~$MeV,
   respectively. We also show the results calculated with the two
values of the gap energy, 1 MeV and 3 MeV. 

As
   one can see in the figures, the values of the partial derivatives have a
   density dependence that makes the derivatives positive and negative. Thus,
   there are subtle cancellations when evaluating the results with
   the potential of Eq. (\ref{eq:pibc}) because, even if the derivatives
act on the
potential
  which depends on the nuclear density, the physical meaning is that one should
   be producing the kaon momentum which is a small quantity. Hence, the
necessary cancellations have to
   appear in the integrals involving the derivatives of the density,
    which measure nucleon momenta, to
   finally account for the small kaon momenta. This is the
   reason why so much emphasis has to be made on the low density limit
because,
   as one can see in Fig.~\ref{fig:b}, the low density limit for the partial derivative
   with respect to $\vec{k}\,^2$ forces a change of sign of the
derivatives with
   respect to $\rho$ which will appear when the $\vec{\nabla}$ operator acts 
 on the $\rho$ dependent functions of the potential. 

\section{P--waves}

   The lowest order meson-baryon Lagrangian
   \cite{leu} can be written as
\begin{eqnarray}
L_1^{(B)} = &&\langle \bar{B} i \gamma^{\mu} \nabla_{\mu} B
\rangle -
M_B \langle \bar{B} B \rangle \nonumber \\
& + &\frac{1}{2} D \langle \bar{B} \gamma^{\mu} \gamma_5 \left\{
u_{\mu},
B \right\} \rangle
+ \frac{1}{2} F \langle \bar{B} \gamma^{\mu} \gamma_5 [u_{\mu},
B]
\rangle \ ,
\label{eq:lag}
\end{eqnarray}
   where the symbol $\langle \rangle$ stands for the trace of the SU(3)
matrices B, involving
   the baryon fields of the octet of the nucleon, and $u_\mu$, involving the
   meson fields of the octet of the pion. The constants $D$ and $F$
are given by $D+F~=~g_A~=1.257$ and $D-F~=~0.33$.
   The term involving the covariant derivative gives rise to a contact
    s--wave term, which is evaluated in \cite{ose}, and a contact 
   p--wave term which we write here in Eq.~(\ref{eq:tpc})  in the
   center of mass (CM) frame of the $\bar{K} N $ state 
\begin{equation}
t^{(p,C)}=-C_{ij} \frac{1}{4f^2} \left(\frac{E_i+M_i}{2 M_i}\right)^{1/2}
\left(\frac{E_j+M_j}{2 M_j}\right)^{1/2}
\left(\frac{1}{2M_i}+\frac{1}{2M_j}\right) 
\vec{\sigma}\vec{k}\,^\prime\, \vec{\sigma}\vec{k} \ ,
\label{eq:tpc}
\end{equation}
     with $M_{i,j}$, $E_{i,j}$, the mass and energy of the initial and final
   baryons, and $C_{ij}$ SU(3) coefficients which are tabulated in
   \cite{ose}. The variables $\vec{k}\,^\prime$ and $\vec{k}$ are the
center
of mass momenta of the outgoing and incoming meson, respectively.
 
   In addition we also have the contribution of the $\Lambda,\Sigma$ and
   $\Sigma^*$ pole diagrams with two vertices of the type $KNY$, with $Y$
being the hyperon, which come from the $D$ and $F$ terms of
Eq.~(\ref{eq:lag}) and that can be easily evaluated
\cite{klin,phi}. 

Unlike the $\Lambda(1405)$ resonance, which is generated
dynamically from the lowest order s--wave chiral Lagrangians and multiple
scattering, the strength
of the p--wave interaction is too small to generate dynamically the
$\Sigma^*(1385)$, which,
in the language of refs.~\cite{derafael,oller}, would then qualify as a
genuine, or preexisting, resonance built up mostly from three-quark
states.
Thus, we follow a phenomenological approach to include the contribution
from $\Sigma^* h$ excitations. The
   $KN\Sigma^*$ vertex is evaluated in \cite{phi} by means of SU(6)
symmetry,
   in analogy to the evaluation of the $\pi \Delta N$ vertex from the 
   $\pi N N$ one. It differs slightly from
   the one used in \cite{klin}, where  SU(3) arguments are used.
    
For the case of $\Lambda$ and $\Sigma$ pole terms  the expressions that we
    get for the p--wave amplitude in the CM are
\begin{equation}
t^{(p,Y)} = 
D_{M^\prime B^\prime Y} 
D_{M B Y} 
\left(1+\frac{k^{\prime\,0}}{2M_{B^\prime}}\right)
\left(1+\frac{k^0}{2M_B}\right)
\frac{1}{\sqrt{s}-M_Y} 
\vec{\sigma}\vec{k}\,^\prime \, \vec{\sigma}\vec{k} \ ; \
Y=\Lambda,\Sigma
\label{eq:tpy}
\end{equation}
    where the subindices $M,B$ (${M}^\prime,{B}^\prime$) stand for the 
initial (final) meson, baryon. The
quantities $D_{MBY}$ are SU(3) coefficients given by
\begin{equation}
D_{MBY} = c_{DY} \sqrt{\frac{20}{3}}\frac{D}{2 f} -
c_{FY} \sqrt{12}\frac{F}{2 f}
\end{equation}
     where the $c_{DY}$, $c_{FY}$ coefficients are given in table
\ref{table1}, and $f$ = 1.15 $f_\pi$ as in ref.~\cite{ose}.
     
     Similarly, in the case of the $\Sigma^*$ pole term the amplitude in the CM 
     is given by 
\begin{equation}
t^{(p,\Sigma^*)} =
D_{M^\prime B^\prime \Sigma^*}
D_{M B \Sigma^*}
\frac{1}{\sqrt{s}-M_{\Sigma^*}}
\vec{S}\vec{k}\,^\prime \, \vec{S}^\dagger\vec{k} \ ,
\end{equation}
     where the $D$ coefficient is given by
\begin{equation}
D_{MB\Sigma^*} = c_{S} \frac{12}{5}\frac{D+F}{2 f} 
\end{equation}
     and $c_S$ is tabulated in table \ref{table1}.

\begin{table}[ht]
\centering
\caption{\small Coefficients for the $K^-NY(K^-N\Sigma^*)$ couplings }
\vspace{0.5cm}
\begin{tabular}{c|ccccc}
        & $c_{D\Lambda}$ & $c_{F\Lambda}$ & $c_{D\Sigma}$
&
$c_{F\Sigma}$ & $c_S$  \\
        \hline
 & & & & & \\
$K^-p$ & $-\displaystyle{\sqrt{\frac{1}{20}}}$ & 
$\displaystyle{\sqrt{\frac{1}{4}}}$ & 
$\displaystyle{\sqrt{\frac{3}{20}}}$ &
$\displaystyle{\sqrt{\frac{1}{12}}}$ & 
$-\displaystyle{\sqrt{\frac{1}{12}}}$ \\
 & & & & & \\
$K^-n$ & $0$ & $0$ & 
$\displaystyle{\sqrt{\frac{3}{10}}}$ &
$\displaystyle{\sqrt{\frac{1}{6}}}$ & 
$-\displaystyle{\sqrt{\frac{1}{6}}}$
\end{tabular}
\label{table1}
\end{table}

     In order to evaluate the p--wave selfenergy we write the amplitudes in
     terms of the kaon and nucleon variables in the frame where the nuclear
     Fermi sea is at rest. The former $(1+k^0/2M_B)\vec{\sigma}\vec {k}$
     vertex of Eq.~(\ref{eq:tpy}), which was given in the CM, becomes now 
\begin{equation}
\vec{\sigma}\vec{k}\, \left(1-\frac{k^0}{2M_Y}\right) - \frac{\vec{\sigma}
\vec{p}}{2}k^0 \left(\frac{1}{M_N}+\frac{1}{M_Y}\right) \ ,
\end{equation}     
where $\vec{p}$ is the nucleon momentum,
     and the combination of the two vertices gives 
\begin{equation}
\vec{\sigma}\vec{k}\, \vec{\sigma}\vec{k}
\left(1-\frac{k^0}{2M_Y}\right)^2
+ \frac{\vec{\sigma} \vec{p}\, \vec{\sigma} \vec{p} }{4}(k^{0})^2
\left(\frac{1}{M_N}+\frac{1}{M_Y}\right)^2
\label{eq:sksk}
\end{equation}     
     where we have already specified the forward direction of the kaons where
     the evaluation of the selfenergy in infinite matter is done.
     
     On the other hand, for the $\Sigma^*$ pole term we have to write the kaon
     momentum in the CM frame and then we have
\begin{equation}
\vec{k}_{CM} \simeq \vec{k}\left(1-\frac{k^0}{M_{\Sigma^*}}\right)-
\frac{k^0}{M_{\Sigma^*}}\vec{p} \ .
\end{equation}
      Thus, the $\vec k_{CM} \vec k_{CM}$ combination becomes
\begin{equation}
\vec{k}_{CM} \cdot \vec{k}_{CM} \simeq
\vec{k}\vec{k}\left(1-\frac{k^0}{M_{\Sigma^*}}\right)^2
+\left(\frac{k^0}{M_{\Sigma^*}}\right)^2\vec{p}\,^2 \ .
\label{eq:kcm}
\end{equation}

\begin{figure}
\begin{center}                                                                
\leavevmode
\makebox[0cm]{
\epsfysize = 160pt
\epsfbox{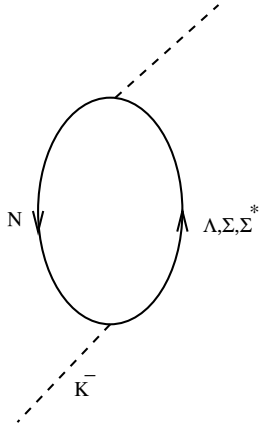}}
\end{center}
\caption{Diagram contributing to p--wave $K^-$-selfenergy via
$K^-N\to\Lambda$,~$K^-N\to\Sigma$ and $K^-N\to\Sigma^*$ processes.}
\label{fig:lind}
\end{figure}

      The evaluation of the selfenergy corresponding to the diagrams
      of Fig.~\ref{fig:lind}  can be written for symmetrical nuclear matter in terms of the 
      Lindhard functions used in \cite{phi} as a p--wave part
\begin{eqnarray}
\Pi^{(p)}_{K^-}(k^0,\vec{k},\rho) &=&
\frac{1}{2} D^2_{K^- p \Lambda} f_\Lambda^2 {\vec k}\,^2
U_\Lambda(k^0,\vec{k},\rho) \nonumber \\
&+& \frac{3}{2} D^2_{K^- p \Sigma^0}
 f_\Sigma^2 {\vec k}\,^2 U_\Sigma(k^0,\vec{k},\rho) \nonumber \\
&+& D^2_{K^- p \Sigma^{*0}}
 f_{\Sigma^*}^2 {\vec k}\,^2 U_{\Sigma^*}(k^0,\vec{k},\rho) \nonumber \\
&-& \frac{3}{2} \frac{1}{4f^2} \frac{1}{M_N} f_c^2 \vec{k}\,^2 \rho
\label{eq:lindp}
\end{eqnarray}
      where the recoil factors $f_i$ are given by
\begin{eqnarray}
f_\Lambda=\left(1-\frac{k^0}{2M_\Lambda}\right) &\ ; \  &
f_\Sigma=\left(1-\frac{k^0}{2M_\Sigma}\right) \nonumber \\
f_{\Sigma^*}=\left(1-\frac{k^0}{M_{\Sigma^*}}\right) &\ ; \ &
f_c=\left(1-\frac{k^0}{M_N + k^0}\right)
\label{eq:recoil}
\end{eqnarray}      
together with an induced piece of s--wave
nature coming from the Fermi motion of the nucleons and the frame 
transformation ($\vec{p}\,^2$ terms in
Eqs.~(\ref{eq:sksk}),(\ref{eq:kcm}))  which is given by
\begin{equation}
\Pi^{(s,ind)}_{K^-}(k^0,\vec{k},\rho) =
\frac{3}{5} k_F^2 (k^{0})^2 (A_p \rho_p + A_n \rho_n)
\label{eq:linds}
\end{equation}
      where $A_N (N=p,n)$ is given by 
\begin{eqnarray}
A_N &=&
\frac{1}{4} \left(\frac{1}{M_N} + \frac{1}{M_\Lambda}\right)^2
\frac{D^2_{K^- N \Lambda}}{\sqrt{s}-M_\Lambda} \nonumber \\
& + & \frac{1}{4} \left(\frac{1}{M_N} + \frac{1}{M_\Sigma}\right)^2
\frac{D^2_{K^- N \Sigma}}{\sqrt{s}-M_\Sigma} \nonumber \\
& + & \left(\frac{1}{M_{\Sigma^*}}\right)^2
\frac{D^2_{K^- N \Sigma^* }}{\sqrt{s}-M_{\Sigma^*}} \nonumber \\
& - & C_{K^-N,K^-N} \frac{1}{4f^2} \frac{1}{M_N} \left(\frac{1}{M_N +
k^0}\right)^2
\end{eqnarray}
      and $C_{K^-p,K^-p}=2$, $C_{K^-n,K^-n}=1$. For the kaonic atom
      case the Lindhard function $U_Y$ is given by:

\begin{equation}
U_Y(m_K,0,\rho)  =  \frac{\rho}{m_K+M_N-M_Y+i{1\over
2}{\Gamma_Y(\sqrt{s}=M_N+m_K)}},\quad Y~=~\Lambda,~\Sigma,~\Sigma^*
\end{equation}
      
Since the momentum in the $\vec{k}\,^2$ factor of Eq.~(\ref{eq:lindp}) comes
from the p--wave amplitudes
and is already in the lab frame, the appropriate combination in finite
nuclei, according to the findings of \cite{seki,juan}, is
$\vec{k}\,^2~U_Y ~\to~-\vec{\nabla} U_Y
\vec{\nabla}$ and $\vec{k}\,^2\,\rho ~\to~-\vec{\nabla} \rho
\vec{\nabla}$. The
familiar ATT term (angular transform term) is here incorporated by means
of the recoil factors $f_i$ given in Eq.~(\ref{eq:recoil}).

    With all these ingredients we solve the KGE, Eq.~(\ref{eq:kg}), and 
     show the results in the next section.  
 
\section{Numerical results}

We solve the KGE of Eq.~(\ref{eq:kg}) with the microscopic
antikaon-nucleus optical potential of ref.~\cite{ram} (neglecting any
isovector effects) plus the several non-local and p--wave terms
discussed in the previous sections. As we already mentioned, for
$V_C(r)$ we use the Coulomb interaction taking exactly the finite-size
distribution of the nucleus and adding to it the vacuum-polarization
corrections~\cite{iz69}.

The numerical solution of the Klein--Gordon equation is done using the 
method in coordinate space of \cite{sal}. The densities used throughout
this work are those compiled in ref.~\cite{baca}.
 However charge (neutron matter) densities do
 not correspond to proton (neutron) ones because of the finite size of
 the proton (neutron). We take that into account following the lines of
 ref.~\cite{juan}.

In the present work we have used a $K^-$ selfenergy calculated in
symmetric nuclear matter. For heavy
nuclei having a neutron excess our
results could be easily improved by weighting the proton and neutron
contributions to the selfenergy in Eq.~(\ref{eq:pik}) with the factors
$Z/(A/2)$ and $N/(A/2)$, respectively. However, for the purpose of the
present work, which is to establish the role of p--wave pieces and
non-localities associated to s--wave, we have not considered necessary to
implement these changes.

\subsection{ P--Wave and non--local effects.}
We solve the KGE using different options to account for the non-local
terms of the $K^-$ selfenergy, $\Pi$.  For each one of these options,
the KGE has been solved for a set of 63 shifts and widths of kaonic
atoms levels\footnote{This set of data is the same used in
\cite{baca}}. Shifts and widths for all models are given in
table~\ref{tab:bind-gap1}~(gap $\Delta$ = 1 MeV) and
table~\ref{tab:bind-gap3}~(gap $\Delta$ = 3 MeV) for a set of selected
typical levels of kaonic atoms across the periodic table. To better
quantify the changes we also give $\chi^2$ per number of data. 
 As can be appreciated in tables the effects on the spectra of both
 the p--wave pieces, discussed
in section 5, and the non-local terms, coming from the dominant
s--wave potential of ref.~\cite{ram} when the zero range approximation
is relaxed, are small when compared to the experimental
uncertainties. 
By comparing $\chi^2/N =3.76$ for row (1), where only the s--wave
local optical potential is considered, with the values of $\chi^2/N$
for rows (2), (3a), (3b), (3c), (3d), (3e) and (4), where different
nonlocal and p--wave effects are added, we see that $\chi^2/N$ change
at most by $\pm$0.7. Then we conclude that the effects of these nonlocal
contributions on the kaonic atoms spectra are smaller than the experimental errors.
Comparison of the results in the two tables shows that
they are also rather independent of the value of the gap. Thus,
despite some theoretical ambiguities in the nonlocal terms, for
practical purposes they can be safely ignored.

The small effect of the p--waves on the shifts and widths found here was
also noted in ref.~\cite{miz}, where the p--wave piece coming from 
$\Sigma^* h$ excitations was included. Their p--wave contribution to the
antikaon selfenergy turned out to be very small compared to the s--wave
one (see Fig.~7 in ref.~\cite{miz}) and made only a ``minor contribution''
to the atomic shifts and widths.

\subsection{ $\Sigma^*$--hole excitation effects on the selfconsistent
  determination of the $s$--wave $K^-$--selfenergy.
\label{sub:sigma}}

\begin{figure}[]
\begin{center}
\leavevmode
\epsfysize=650pt
\makebox[0cm]{\epsfbox{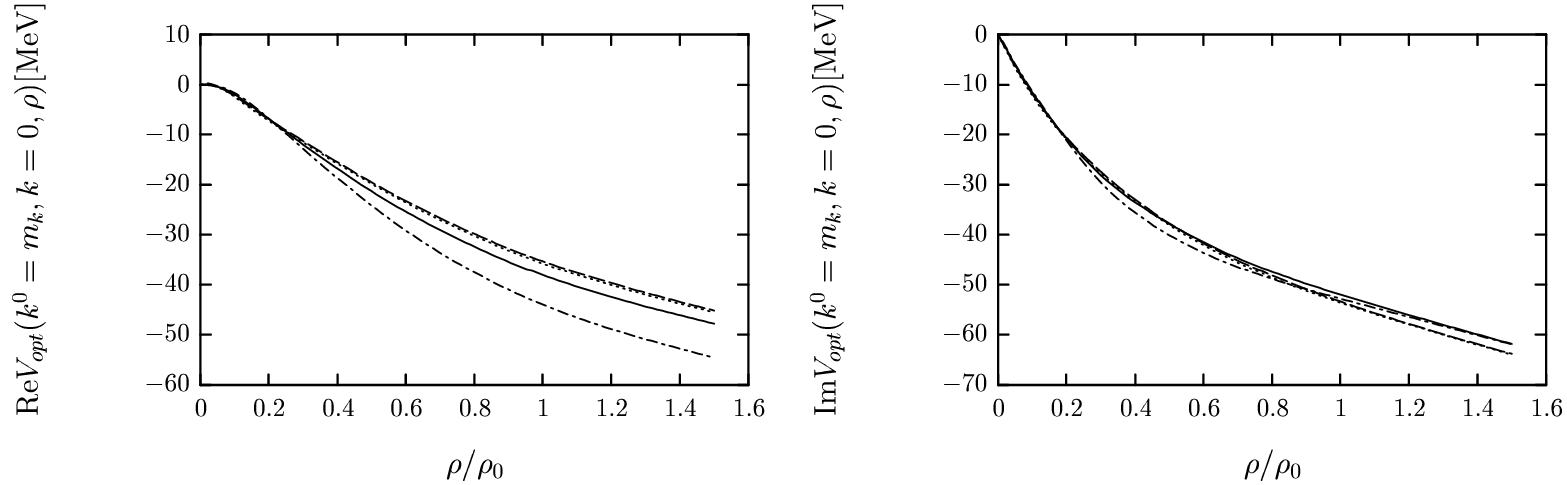}}
\end{center}
\vspace*{-14.4cm}
\caption[pepe]{{ The real and imaginary parts of the $K^-$ optical
    potential, $V_{opt}=\Pi/(2 m_K)$, at energy
 $k^0=m_{K^-}$ and momentum $k=0$ versus $\rho/\rho_0$, are depicted
 in the figures. 
The selfenergy  $\Pi$ is evaluated from Eq.~(\protect{\ref{eq:pik}}). 
 
} The solid  lines have been obtained using no energy 
gap ($\Delta~=~0$), the dashed ones with $\Delta~=~1$~MeV and the
dotted ones with $\Delta~=~3$~MeV. The dot-dashed lines correspond to
the $\Delta=0$ and omitting the $\Sigma^*$-hole propagation of the
$K^-$ in the Bethe--Salpeter equations.}
\label{fig:pigap}
\end{figure}

In this subsection we report on the
role played by the $\Sigma^* h-$excitation when it is included in the
self-consistent calculation of the $s$--wave $K^-$ potential.  In
ref.~\cite{ram}, only the $\Lambda h-$ and $\Sigma h-$$p$--wave terms
where included in the selfconsistent calculation of the dominant
$s$--wave antikaon-nucleus potential. The results obtained with this
potential were reported in refs.~\cite{zaki} and~\cite{baca}, and have
served us here, see row (1) in tables~\ref{tab:bind-gap1}
and~\ref{tab:bind-gap3}, as a reference to evaluate the effect of the
non-local and p--wave terms studied in this paper. In a latter work,
ref.~\cite{phi}, in addition to the $\Lambda h-$ and $\Sigma h-$
$p$--wave terms, the $\Sigma^*$--hole excitation term of
Fig.~\ref{fig:lind} was included in the selfconsistent evaluation of
$s$--wave $K^--$selfenergy, as well.  With this new potential, a
better description of the kaonic atom data is achieved. Indeed, this
improved s--wave potential provides\footnote{This value is obtained with a
  zero gap energy, and only very small changes are found for
  non-vanishing values of the gap energy.} a $\chi^2/N = 2.89$ to be
compared to the value of 3.76 provided by the original potential of
ref.~\cite{ram}. In Fig.~\ref{fig:pigap} we show the s--wave $K^-$--optical
potential  both, with and
without $\Sigma^* h-$ excitation in the selfconsistent evaluation.
Differences between both potentials are moderately small
for the low densities relevant in kaonic atoms, but have a significant
density dependence. At low densities, the inclusion of the
$\Sigma^*h-$term leads to smaller, in absolute value, values for both
the imaginary and real parts of the potential. Thus, there are two
competing effects: reduction of the attraction because of a smaller
real part and a reduction of the repulsion (increase of the
attraction) because of a smaller imaginary part. It seems, that the
latter effect is bigger than the former one and the resulting effect
is an increase of the attraction which leads to a better description
of the data in agreement with the findings of ref.~\cite{baca}. Thus,
the new potential provides bigger widths and smaller, in absolute
value, shifts than the one of ref.~\cite{ram}.

In the row labeled (1)$_{\Sigma^*}$ of table~\ref{tab:bind-gap1}, we
present  the results
for selected kaonic atom levels which are obtained by using only the
s--wave optical
potential which includes the $\Sigma^*-h$ excitation in the selfconsistent
evaluation of the s--wave $K^-$ selfenergy, with a vanishing
value of the gap. By comparing this row with row (1) of the same
table, we see that the inclusion of the $\Sigma^*-h$ excitation in the
selfconsistent evaluation of the s--wave $K^-$ selfenergy improves 
the global agreement between the
theoretically predicted values and the 
the empirical ones. We should note that
unlike the nonlocal effects which depend on the prescription, the inclusion of
the $\Sigma^*-h$ correction in the $s$--wave $K^-$ selfenergy, although it has a
moderate effect, is well defined. On the other
hand, considering again only the s--wave $K^-$ selfenergy, the effect of 
including non vanishing gap values of about $\Delta=1\sim
3$~MeV  has a very small
effect on the shifts and widths of the known kaonic levels. This small
effect can be quantified by observing that, for the considered set of
data, the values 
$\chi^2/N=2.89$ for $\Delta=0$, 
$\chi^2/N=2.83$ for $\Delta=1$~MeV and
$\chi^2/N=2.94$ for $\Delta=3$~MeV, which we obtain using only the s--wave 
$K^-$ selfenergy, are very close.

{\small
\begin{table}
\begin{center}
\begin{tabular}{|l||r|rr|rr|rr|rr|rr|}\hline\tstrut
 \  &$\chi^2/N$
&\multicolumn{2}{|c|} {$_5^{10}$B}
&\multicolumn{2}{|c|} {$_{13}^{27}$Al}
&\multicolumn{2}{|c|} {$_{29}^{63}$Cu}
&\multicolumn{2}{|c|} {$_{48}^{112}$Cd}
&\multicolumn{2}{|c|} {$_{92}^{238}$U}
\\
 &  
 & $-\epsilon_{2p}$ & $\Gamma_{2p}$
 & $-\epsilon_{3d}$ & $\Gamma_{3d}$
 & $-\epsilon_{4f}$ & $\Gamma_{4f}$
 & $-\epsilon_{5g}$ & $\Gamma_{5g}$
 & $-\epsilon_{7i}$ & $\Gamma_{7i}$
 \\\hline\hline\tstrut 
(1) & 3.76
& 217 &  551
& 109 &  368
& 384 & 1121
& 528 & 1437
& 330 & 1090
\\\hline
(2)   & 4.00
& 213 &  542
& 110 &  362
& 392 & 1110
& 543 & 1420
& 350 & 1076
\\\hline
(3a) & 3.20
& 211 &  565
& 102 &  397
& 361 & 1229
& 494 & 1588 
& 302 & 1291
\\\hline
(3b) & 4.00
& 234 &  564 
& 118 &  383
& 415 & 1172
& 568 & 1515
& 357 & 1196
\\\hline
(3c) & 4.01
& 234 &  564
& 118 &  383
& 415 & 1173
& 569 & 1515
& 358 & 1197
\\\hline
(3d) & 4.02 
& 233 &  562
& 118 &  382
& 414 & 1170
& 568 & 1511
& 356 & 1194
\\\hline
(3e) & 3.38
& 219 &  568
& 110 &  383
& 391 & 1182
& 538 & 1528
& 336 & 1203
\\\hline
(4) & 3.69
& 217 &  552
& 110 &  371
& 388 & 1141
& 534 & 1465
& 337 & 1166
\\\hline
(1)$_{\Sigma^*}$ & 2.89
& 208 &  575
& 105 &  398
& 373 & 1219
& 512 & 1550
& 320 & 1201
\\\hline\hline
Exp & - 
& 208 &  810
&  80 &  443
& 370 & 1370
& 400 & 2010
& 260 & 1500
\\  
   &
& $\pm$35 & $\pm$100
& $\pm$13 & $\pm$22
& $\pm$47 & $\pm$170
& $\pm$100 & $\pm$440
& $\pm$400 & $\pm$750
\\\hline
\end{tabular}
\end{center}

\vspace*{-.25cm}
\caption[pepe]{ \small Widths and shifts of representative  kaonic
atom levels in 
eV obtained from different potentials, taking always the energy gap
equal to $\Delta = 1 $~MeV. Different cases correspond to
the local potential of ref.~\protect\cite{ram} (for
a  comparison in row (1) we give the results obtained with
$\Pi(m_K,0,\rho)$, first term on the r.h.s. of
Eq.~(\ref{eq:pibc})) . Rows (2) to (4)
correspond to
different additions to this dominant piece:

\hspace*{0.4cm}
(2) Only p--wave Lindhard function 
non-local effects from ~Eq.~(\ref{eq:lindp}) and the local
induced term of Eq.~(\ref{eq:linds}) are added. 

\hspace*{0.4cm}
(3) Only $b\,
\vec{q}\,^2$ non-local effects, see Eq.~(\ref{eq:pibc}), are included
using different ways: (3a) 
$-\vec\nabla b\vec\nabla$, (3b) $-\vec\nabla b\vec\nabla -0.5 (\Delta b)$, (3c)
$-b\vec\nabla^2$, (3d) $b~[(\omega-V_C)^2-\mu^2-\Pi]$, (3e) ${b\,{\rm
Re}[(\omega-V_C)^2-\mu^2-\Pi]}$.

\hspace*{0.4cm}
(4) Only $c\, (\omega-\mu)$ ``non-local''
effects are added, see Eq.~(\ref{eq:pibc}).

The results of row $(1)_{\Sigma^*}$ are obtained from a purely
$s$--wave optical potential, like in row (1), but including 
$\Sigma^*-h$ excitations as described in subsection \ref{sub:sigma}. 
For each potential, $\chi^2$ per number of data
$N$ is shown. We use a set of 63 shifts and widths previously used in
ref.~\protect\cite{baca}.
\vspace*{0.75cm}    }
\label{tab:bind-gap1}
\end{table}
}


{\small
\begin{table}
\begin{center}
\begin{tabular}{|l||r|rr|rr|rr|rr|rr|}\hline\tstrut
 \  &$\chi^2/N$
&\multicolumn{2}{|c|} {$_5^{10}$B}
&\multicolumn{2}{|c|} {$_{13}^{27}$Al}
&\multicolumn{2}{|c|} {$_{29}^{63}$Cu}
&\multicolumn{2}{|c|} {$_{48}^{112}$Cd}
&\multicolumn{2}{|c|} {$_{92}^{238}$U}
\\
 &  
 & $-\epsilon_{2p}$ & $\Gamma_{2p}$
 & $-\epsilon_{3d}$ & $\Gamma_{3d}$
 & $-\epsilon_{4f}$ & $\Gamma_{4f}$
 & $-\epsilon_{5g}$ & $\Gamma_{5g}$
 & $-\epsilon_{7i}$ & $\Gamma_{7i}$
 \\\hline\hline\tstrut 
(1) & 3.76
& 217 &  551
& 109 &  368
& 384 & 1121
& 528 & 1437
& 330 & 1090
\\\hline
(2)   & 4.00
& 213 &  542
& 110 &  362
& 392 & 1110
& 543 & 1420
& 350 & 1076
\\\hline
(3a) & 3.03
& 207 &  567
&  97 &  402
& 343 & 1254
& 469 & 1623 
& 272 & 1335
\\\hline
(3b) & 3.96
& 235 &  566 
& 119 &  386
& 416 & 1187
& 569 & 1539
& 353 & 1225
\\\hline
(3c) & 3.95
& 235 &  567
& 119 &  386
& 416 & 1188
& 569 & 1540
& 353 & 1225
\\\hline
(3d) & 3.95 
& 235 &  565
& 119 &  386
& 415 & 1184
& 567 & 1534
& 351 & 1220
\\\hline
(3e) & 3.25
& 218 &  570
& 109 &  386
& 387 & 1193
& 531 & 1548
& 325 & 1227
\\\hline
(4) & 3.65
& 217 &  552
& 109 &  371
& 386 & 1142
& 531 & 1479
& 330 & 1169
\\\hline\hline
Exp & - 
& 208 &  810
&  80 &  443
& 370 & 1370
& 400 & 2010
& 260 & 1500
\\  
   &
& $\pm$35 & $\pm$100
& $\pm$13 & $\pm$22
& $\pm$47 & $\pm$170
& $\pm$100 & $\pm$440
& $\pm$400 & $\pm$750
\\\hline
\end{tabular}
\end{center} 
\vspace*{-.25cm}
\caption[pepe]{
\small Same as in table~\protect\ref{tab:bind-gap1}
 but for an energy gap of $\Delta=3 $~MeV}
\label{tab:bind-gap3}
\end{table}
}

\section{Conclusion}
   We have concentrated on the evaluation of corrections to the $K^-$ nucleus 
 optical potential originating from the momentum and energy dependence of the
 s--wave selfenergy, previously developed, plus corrections to this potential
 originating from the p--wave part of the elementary $K^-N$ amplitudes. Some of
 the corrections lead to nonlocal terms in finite nuclei, and the fact that the
 calculations of the selfenergy are done previously using local approximations,
 like the assumption of a local Fermi sea, introduces some ambiguities
 in the choice of the form of the nonlocal potential. The use of several
 alternatives serves us to quantify the amount of intrinsic uncertainty in
 our approach. 

Fortunately, the corrections turn out to be small, smaller than
 the experimental errors, which gives support to the results obtained with
 only an s--wave optical potential, which has been the rule in the studies of
 kaonic atoms. We also make some choices of equivalent, energy dependent,
 potentials and find also small corrections. 
 We have calculated widths and shifts for
 a large set of kaonic atoms. Although fits of better quality than the present
 results can be obtained, the potential has the merit of being a free parameter
 theoretical one which gives a fair reproduction of the data. 

We have made a thorough study of the optical potential
around threshold and have shown the mathematical problems that one faces at
very low densities. They are related to the cusp in the elementary $K^- N
\to K^- N$ scattering amplitude at threshold, where some of the derivatives
become infinite. The peculiar low density behaviour found in \cite{flor} is
tied to these problems and a natural physical solution for them is found here
by considering the finite excitation energy of real nuclei.

   As with respect to the use of empirical potentials to analyze the kaonic atom
 data, our results endorse the approaches which are based exclusively on an 
 s--wave optical potential, once we see that all nonlocal terms generated lead 
 to corrections which are smaller than present experimental errors.

Another relevant finding of this paper is,  that using a purely
theoretical optical potential, a quite satisfactory  description of
the kaonic atom data is achieved. For 63 data, we find $\chi^2/
N = 2.89$, using the s--wave local theoretical potential of ref.~\cite{phi} which contains
$\Sigma^*-h$ excitations in the selfconsistent calculation of the
s--wave $K¯$ selfenergy.

\section{Acknowledgment}
We would like to thank A. Gal for discussions.  This work is partly
supported by DGICYT contracts PB96-0753, PB98-1247 and PB98-1367.  We
would also like to acknowledge support from the EU TMR Network
Eurodaphne, contract no.  ERB\-FMRX-CT98-0169, and from the
Generalitat de Catalunya and Junta de Andaluc\'{\i}a  under grants
2000SGR-24 and FQM 0225 respectively.
\renewcommand{\theequation}{\Alph{section}.\arabic{equation}} 
\setcounter{section}{1}
\setcounter{equation}{0}                                         
\section*{Appendix: the Lindhard function with a gap in the particle-hole
excitation energy}

We define the Lindhard function for the forward going particle-hole excitation
as
\begin{equation}
U(q^0,q,\Delta,\rho)=4\int \frac{d^4 k}{(2\pi)^3}
\frac{n(\vec{k}\,,\rho)[1-n(\vec{k}+\vec{q}\,,\rho)]}{q^0+\varepsilon(\vec{k}\,) -
\varepsilon(\vec{k}+\vec{q}\,) - \Delta + i\eta} \ .
\label{eq:ugap}
\end{equation}
where the energy gap $\Delta$ separates the occupied and unoccupied
nucleon states. Through the paper in some situations we have
referenced to $U(q^0,q,\rho)$ which is obtained from
Eq.~(\ref{eq:ugap}) above, 
setting the gap to zero.
We find for $x\leq 2$:
\begin{eqnarray}
{\rm Re}\, U(q^0,q,\Delta,\rho) &=& -\frac{2 M k_F}{\pi^2}
\frac{1}{2x}\left\{\frac{x}{2}-\frac{\nu-\delta}{4} + 
\frac{\nu-\delta}{2}\, \ln \left| \frac{\nu-\delta + x^2 -2x}{\nu -\delta} \right|
\right. \nonumber \\
&+ &
\left. \frac{1}{2} \left[ 1 -
\frac{1}{4}\left(\frac{\nu -\delta}{x}-x\right)^2 \right] 
\ln \left| \frac{\nu -\delta -x^2-2x}{\nu -\delta +x^2-2x} \right| \right\} \ ,
\end{eqnarray}
and for $x\geq 2$:
\begin{equation}
{\rm Re}\, U(q^0,q,\Delta,\rho) = -\frac{2 M k_F}{\pi^2}
\frac{1}{2x}\left\{\frac{x}{2}-\frac{\nu-\delta}{2x} + 
\frac{1}{2}\left[ 1 -
\frac{1}{4}\left(\frac{\nu -\delta}{x}-x\right)^2 \right] 
\ln \left| \frac{\nu -\delta -x^2-2x}{\nu -\delta -x^2+2x} \right| \right\} \ ,
\end{equation}
where
\begin{equation}
\nu = \frac{2Mq^0}{k^2_F} \ , ~~~x=\frac{q}{k_F}\ , 
~~~\delta=\frac{2M\Delta}{k^2_F}\ ,
~~~\rho = \frac{2 k_F^3}{3\pi^2} \ .
\end{equation}

For the imaginary part we find:
\begin{equation}
{\rm Im}\, U(q^0,q,\Delta,\rho) = \left\{
\begin{array}{ll}
{\rm Im}\, \overline{U}(q^0-\Delta,q,\rho) & ~~~{\rm for}\  q^0 > \Delta \\ 
0 & ~~~{\rm for}\ q^0 < \Delta
\end{array}
\right.
\end{equation}
where
\begin{equation}
{\rm Im}\, \overline{U}(q^0,q,\rho)=-\frac{3}{4}\pi\rho\frac{M}{q k_F}
\left[ (1-z^2) \theta(1-\mid z \mid) - (1-z^{\prime\,2}) \theta(1 - \mid z^\prime
\mid) \right] \frac{q^0}{\mid q^0 \mid} \ ,
\end{equation}
with
\begin{equation}
z=\frac{M}{q k_F} \left(q^0-\frac{q^2}{2M}\right) , ~~
z^\prime=\frac{M}{q k_F} \left(-q^0-\frac{q^2}{2M}\right) \ .
\end{equation}


\end{document}